\documentclass[useAMS,usenatbib]{mn2e}
\usepackage{natbib}
\usepackage{apjfonts}
\usepackage{amsmath}
\usepackage{epsfig}
\usepackage{graphicx}
\usepackage{amssymb}
\usepackage{aas_macros}
\usepackage{color}

\newcommand{\kms}{\,{\rm km \, s^{-1}}}
\newcommand{\oversim}[2]{\protect{\mbox{\lower0.5ex\vbox{%
   \baselineskip=0pt\lineskip=0.2ex
   \ialign{$\mathsurround=0pt #1\hfil##\hfil$\crcr#2\crcr\sim\crcr}}}}} 

\catcode`\"=\active\let"=\"  

\def\3{{\ss} }

\def\c12{{1\over 2}}

\def\d{{\rm d}}   
   
\def\plusplus{\raise 0.3ex\hbox{${\scriptstyle ++}$}{}}

\bibliography{biblio}

\begin{document}   

\title[Dynamical invariants and diffusion of substructures]{Dynamical invariants and diffusion of merger substructures in time-dependent gravitational potentials} 
\author[Jorge Pe\~{n}arrubia]{Jorge Pe\~{n}arrubia$^{1,2}$\thanks{jorpega@roe.ac.uk}\\
$^1$Institute for Astronomy, University of Edinburgh, Royal Observatory, Blackford Hill, Edinburgh EH9 3HJ, UK\\
$^2$Ram\'on y Cajal Fellow, Instituto de Astrof\'isica de Andaluc\'ia-CSIC, Glorieta de la Astronom\'ia, 18008, Granada, Spain }

\maketitle

\begin{abstract} 
This paper explores a mathematical technique for deriving dynamical invariants (i.e. constants of motion) in time-dependent gravitational potentials. The method relies on the construction of a canonical transformation that removes the explicit time-dependence from the Hamiltonian of the system. By referring the phase-space locations of particles to a coordinate frame in which the potential remains `static' the dynamical effects introduced by the time evolution vanish. It follows that dynamical invariants correspond to the integrals of motion for the static potential expressed in the transformed coordinates. The main difficulty of the method reduces to solving the differential equations that define the canonical transformation, which are typically coupled with the equations of motion. We discuss a few examples where both sets of equations can be exactly de-coupled, and cases that require approximations. 
The construction of dynamical invariants has far-reaching applications. These quantities allow us, for example, to describe the evolution of (statistical) microcanonical ensembles in time-dependent gravitational potentials without relying on ergodicity or probability assumptions. As an illustration, we follow the evolution of dynamical fossils in galaxies that build up mass hierarchically. It is shown that the growth of the host potential tends to efface tidal substructures in the integral-of-motion space through an orbital diffusion process. The inexorable cycle of deposition, and progressive dissolution, of tidal clumps naturally leads to the formation of a `smooth' stellar halo.

\end{abstract}   

\section{Introduction} \label{sec:intro}
The description of dynamical systems out of equilibrium remains an outstanding problem in Physics and Astronomy. Hamilton was among the first to attack it via the construction of `perturbed' Hamiltonians for systems that are close to an equilibrium state 
\begin{equation}\label{eq:hamilpert}
H({\bf q},{\bf p},t)\approx H_0({\bf q},{\bf p})+\epsilon H_1({\bf q},{\bf p},t)+ ...+ \epsilon^k H_k({\bf q},{\bf p},t);
\end{equation}
 where $({\bf q}, {\bf p})$ are the coordinates of a particle in configuration space, $t$ is the time and $\epsilon\ll 1$. In perturbative methods solutions to the equations of motion are calculated iteratively from low to high order. Unfortunately, the trajectories of particles through phase space can be rarely expressed analytically, limiting the applicability of this method to the simplest of cases (e.g. Binney \& Tremaine 2008). An improved perturbation theory is obtained by expressing the Hamiltonian in terms of action-angle variables. This technique is particularly attractive for Hamiltonians that are completely separable. In these systems the actions (${\bf J})$ associated with the time-independent term ($H_0$) are conserved along the path of a particle motion and define the surface of a torus in phase-space, while angle variables vary linearly with time and provide the coordinates of a particle on the torus. Yet, this approach has its own drawbacks. For example, the analytical expression of action variables is only possible for a few cases of astronomical interest, namely the Keplerian, harmonic and isochronic potentials (e.g. Binney 2010, 2012a,b). Furthermore, in time-varying potentials actions do not remain constant but oscillate with an amplitude $|\Delta J/J_{0}|\propto \epsilon T_0$, where $\epsilon\equiv \dot \Phi/\Phi_0$ is the time-derivative of the gravitational potential and $T_0$ is the period associated with the motion on the torus. When averaged over several orbital periods, actions are conserved at order $\mathcal{O}(\epsilon T_0)^2$. Thus, identifying $\epsilon$ with the growth rate of the potential in Equation~(\ref{eq:hamilpert}) shows that perturbation theories can only be applied to systems that evolve in an adiabatic regime.


Perturbative methods are {\it deterministic}, i.e. they rest upon solutions to the dynamical equations of motion. 
A fundamentally different description of many-body systems approaching dynamical equilibrium is provided by the construction of {\it statistical} ensembles, in which trajectories of particles are replaced by a probability distribution of finding sets of particles in a given phase-space volume. 
Although mechanical statistics successfully describe the gross evolution of systems evolving under a rapidly-varying gravitational field, such as those driving `violent relaxation' processes (Lynden-Bell 1967; Tremaine et al. 1986; Ponzten \& Governato 2013), it was early realized that deterministic and statistical descriptions of systems subject to long-range forces do not lead to the same physical behaviour (e.g. Hertel \& Thirring 1971; see Padamanabhan 1990 for a review). Indeed, gravitationally bound objects have negative specific heat (Antonov 1961; Lynden-Bell \& Lynden-Bell 1977; Padmanabhan 1989; see Lynden-Bell 1999 for a review), a quantity that must be positive definite in ensembles where the energy of individual particles is allowed to fluctuate probabilistically about a time-average value, as in the canonical and grand canonical distributions. 
As a consequence gravitating systems must be described by microcanonical ensembles, in which the energy of individual particles is kept fixed. In order to derive the microcanonical distribution it is commonly assumed that all individual microstates on a given energy surface in phase-space are equally probable (the so-called ergodic hypothesis). This assumption guarantees that the time-averaged properties of microcanonical ensembles can be directly derived from a phase-space average over all possible microstates. However, the equivalence between time average and average over ensembles only arises when the system can visit all the possible microstates, many times, during a long period of time. In the case of ensembles out of dynamical equilibrium, which contain transient microstates by definition, the ergodic assumption may lead to a biased description of the system.

This paper explores a mathematical tool for deriving the dynamical evolution of microcanonical ensembles in time-dependent gravitational potentials which does not rely on ergodicity or probability assumptions. Instead, a canonical transformation is constructed (\S\ref{sec:dyninv}) that removes the explicit time-dependence from the Hamiltonian of the system. By referring the phase-space locations of the particle ensemble to a coordinate frame in which the potential remains `static', the dynamical effects introduced by the time evolution disappear. It follows that dynamical invariants (i.e. quantities that are conserved along the phase-space path of a particle) can be straightforwardly constructed by expressing the integrals of motion for the static potential in the transformed coordinates. The main difficulty of this technique reduces to solving the differential equations that define the canonical transformation, which are typically coupled with those that define the trajectory of particles through phase space. Section~\ref{sec:examples} discusses a few examples of astronomical interest where both sets of equations can be de-coupled.

The construction of invariants allows us to describe the macroscopical (statistical) properties of large ensembles of gravitating particles through a simple time averaging of microscopic (deterministic) equations. As an illustration, we study the thermodynamics of cold tidal substructures orbiting in a time-dependent potential in Section~\ref{sec:appl}. In particular, we follow the evolution of entropy, temperature and specific heat, and compare the results against those derived from mechanical statistics. 

In Sections~\ref{sec:scatt} and~\ref{sec:entroev} we use dynamical invariants to describe the evolution of dynamical fossils in galaxies that build up mass hierarchically. This is a timely issue given that Gaia (Perryman et al. 2001) is expected to uncover a large number of accreted substructures in the integral-of-motion space (Helmi \& de Zeeuw 2000; Brown et al. 2005; G\'omez et al. 2010; Sharma et al. 2011; Mateu et al. 2011; although see Valluri et al. 2012).  Given that integrals are not conserved quantities in hierarchical models of galaxy formation, we put special emphasis on understanding the diffusion of tidal substructures in the integral-of-motion space. Our analytical results are illustrated by means of restricted $N$-body models in Section~\ref{sec:nbody}. Section~\ref{sec:detect} discusses the detectability of tidal substructures. The conclusions are laid out in Section~\ref{sec:conc}.

\section{Dynamical invariants}\label{sec:dyninv}
The method for constructing dynamical invariants proposed below is conceptually simple. 
 The goal is to find a coordinate transformation in which the explicit time-dependence of the potential vanishes, so that the integrals of motion in the transformed coordinates become the desired dynamical invariants. For simplicity, our calculations are derived in the mean-field limit, thus ignoring the granularity of $N$-body systems.

\subsection{Newtonian formulation}\label{sec:newton}
Considering a universe in which Newton's constant $G$ decreases, Lynden-Bell (1982) found a coordinate transformation that recovers standard equations of motion (see \S\ref{sec:dirac}). This Section generalizes Lynden-Bell's arguments for any system with a time-dependent gravitational potential. 

Let us first write the equations of motion of a particle subject to a time-varying force ${\bf F}({\bf r},t)$ as
\begin{equation}
\label{eq:eqmot1}
\ddot{\bf r}= {\bf F}({\bf r},t).
\end{equation}
The simplest distance transformation that one can introduce is the following
\begin{equation}
\label{eq:transf}
{\bf r}={\bf r}'R(t);
\end{equation}
so that the left-hand term in Equation~(\ref{eq:eqmot1}) becomes
\begin{equation}
\label{eq:accel}
\ddot{\bf r}=R\ddot {\bf r}'+2\dot R\dot{\bf r}'+\ddot R{\bf r}'.
\end{equation}
Now we define a new time coordinate 
\begin{equation}
\label{eq:time}
\d \tau= f(t)\d t;
\end{equation}
so that Equation~(\ref{eq:accel}) becomes
\begin{equation}
\label{eq:accel2}
\ddot{\bf r}=R f^2 \frac{\d^2 {\bf r}'}{d \tau^2} + (2\dot R f +R \dot f)\frac{\d {\bf r}'}{\d \tau}+ \ddot R {\bf r}'.
\end{equation}
For $f=R^{-2}$ the velocity term vanishes, and the equation of motion becomes
\begin{equation}
\label{eq:eqmot_trans}
\frac{\d^2 {\bf r}'}{d \tau^2}+\ddot R R^3 {\bf r}'- R^3 {\bf F}(R{\bf r}',t)=0.
\end{equation}
We are still free to choose the time-dependent scaling function $R(t)$. One would like to use this freedom to identify Equation~(\ref{eq:eqmot_trans}) with the equations of motion in a time-independent potential, i.e.
\begin{equation}
\label{eq:eqmot_tind}
\frac{\d^2 {\bf r}'}{d \tau^2}-{\bf F}'[{\bf r}']=0.
\end{equation}
From Equations~(\ref{eq:eqmot_trans}) and~(\ref{eq:eqmot_tind}) the scaling factor must be a solution of the following differential equation
\begin{equation}
\label{eq:emar_R}
\ddot R R^3 {\bf r}'- R^3 {\bf F}(R{\bf r}',t)= -{\bf F}'[{\bf r}'].
\end{equation}
Clearly, if ${\bf F}$ is a conservative force (${\bf \nabla}\times{\bf F}=0$) then ${\bf F}'$ is also conservative. Therefore, it is possible to define a time-independent scalar potential $\Phi'=-\int F'\d r'$, so that in the transformed coordinates the energy ($I$) becomes an exact dynamical invariant (i.e. a constant of motion)
 \begin{equation}
\label{eq:inv}
I = \frac{1}{2}\bigg(\frac{\d {\bf r}'}{\d \tau}\bigg)^2+\Phi'({\bf r}')= \frac{1}{2}(R \dot {\bf r}- \dot R {\bf r})^2 + \frac{1}{2}\ddot R R r^2 + R^2\Phi({\bf r},t);
\end{equation}
where $R(t)$ is a solution of~(\ref{eq:emar_R}) and $\Phi({\bf r},t)=-R\int {\bf F}(R{\bf r}',t)\d{\bf r}'$. 

Note also that the angular momentum ${\bf L}$ remains invariant under the transformation $\d{\bf r}'/\d \tau =R\dot {\bf r} - \dot R {\bf r}$, i.e.
\begin{equation}
\label{eq:angmom}
{\bf L}={\bf r}'\times \frac{\d {\bf r}' }{\d \tau}={\bf r}\times \dot {\bf r}.
\end{equation}
In general it is straightforward to show that all classical integrals of the equations of motion of a Newtonian potential $\Phi'({\bf r}')$ reappear as the constants of motion of $\Phi({\bf r},t)$. This result also applies to integrals derived numerically, although obtaining these quantities is usually difficult (e.g. Bienaym\'e \& Trevon 2013 and references therein).

\subsection{Hamiltonian formulation}\label{sec:hamilton}
Statistical mechanics provides a powerful tool in order to understand the physical behaviour of gravitating systems composed of many particles. For such analysis it is useful to generalize the results obtained in \S\ref{sec:newton} using the Hamiltonian formalism. The Hamiltonian of a system with $\nu-$degrees of freedom can be written as
\begin{equation}
\label{eq:ham1}
H=\sum_{i=1}^\nu \frac{1}{2}p_i^2 + \Phi({\bf q},t);
\end{equation}
where $(q_1,...,q_\nu; p_1,...,p_\nu)$ are the coordinates of a particle in configuration space. The equations of motion are
\begin{subequations}
\begin{align}
\dot q_i=p_i \label{eq:h_dotq}\\
\dot p_i=-\frac{\partial \Phi}{\partial q_i}\label{eq:h_dotp}.
\end{align}
\end{subequations}

The fact that the dynamical invariants derived in \S\ref{sec:newton} correspond to an energy in a new coordinate system suggests that there must exist a canonical transformation $(q_i,p_i)\rightarrow (q_i',p_i')$ that removes the explicit time dependence from the Hamiltonian. To find such a transformation we first consider an intermediate Hamiltonian $\hat H$ and a time-dependent generating function $Q$, so that $\hat H({\bf q}',{\bf p}',t)=H({\bf q}',{\bf p}',t)+\partial Q({\bf q}',{\bf p}',t)/\partial t$ (see also Lewis \& Leach 1982; Struckmeier \& Riedel 2001). 

Following Equation~(\ref{eq:inv}), the goal is to find a generating function that yields the transformations $p_i'=R p_i -\dot R q_i$, and $q_i=R q_i'$. It is straightforward to show that the function
\begin{equation}
\label{eq:ham1}
Q({\bf q}',{\bf p},t)=\sum_{i=1}^\nu \bigg[\frac{1}{2}R \dot R q_{i}'^2 - R p_i q_i'\bigg];
\end{equation}
is the desired one given that 
\begin{subequations}\label{eq:trans_h}
\begin{align}
q_i=-\frac{\partial Q}{\partial p_i} = R q_i' \label{eq:trans_q}\\
p_i'=-\frac{\partial Q}{\partial q_i'} = R p_i- R\dot R q_i'=R p_i- \dot R q_i. \label{eq:trans_p}
\end{align}
\end{subequations}

Before we calculate the new Hamiltonian $\hat H=H+\partial Q/\partial t$ recall that our goal is to find a dynamical invariant that is conserved {\it along the phase-space path of a particle motion}, i.e. the subset of the $6N$ dimensional phase space on which the equations of motion~(\ref{eq:h_dotq}) and~(\ref{eq:h_dotp}) are fulfilled. This means that along the phase-space path all terms in Equation~(\ref{eq:ham1}) that depend on the particle trajectory are functions of $t$ only, so that the canonical transformation yields the following Hamiltonian
\begin{eqnarray}
\label{eq:ham2}
\hat H({\bf q}',{\bf p}',t)=\frac{1}{R^2}\bigg[\frac{1}{2}\sum_{i=1}^\nu p_i'^2 +\frac{1}{2} \ddot R R^3 q_i'^2 + R^2\Phi(R{\bf q}',t)\bigg]= \\ \nonumber
\frac{1}{R^2}\bigg[\frac{1}{2} \sum_{i=1}^\nu p_i'^2 +\hat \Phi({\bf q}',t)\bigg];
\end{eqnarray}
where $\hat \Phi({\bf q}',t)\equiv\ddot R R^3 \sum_i (1/2) q_i'^2 + R^2\Phi(R{\bf q}',t)$.

The time still appears explicitly in Equation~(\ref{eq:ham2}) through $R(t)$ and $\hat \Phi({\bf q}',t)$. This dependence can be eliminated in two steps. First, through a re-scaling of the time coordinate. Choosing $f=R^{-2}$ in Equation~(\ref{eq:time}) yields the equations of motion $\d q_i'/\d \tau=R^{2}\partial \hat H/\partial p_i'=p_i'$ and $\d p_i'/\d \tau=-R^{2}\partial \hat H/\partial q_i'$. Hence, the Hamiltonian $H'=R^2\hat H= \sum_i\frac{1}{2}p_i'^2 +\hat \Phi({\bf q}',t)$ has a time-dependence through $\hat \Phi$ only.

Next, as in \S\ref{sec:newton}, the freedom in the time-dependent scaling function $R(t)$ can be used to remove the explicit time-dependence from the potential $\hat \Phi$ so that
\begin{equation}
\label{eq:hamR}
\ddot R R^3 \sum_{i=1}^\nu \frac{1}{2} q_i'^2 + R^2\Phi(R{\bf q}',t) = \Phi'({\bf q}'). 
\end{equation}
Note that Equation~(\ref{eq:emar_R}) is recovered by differentiating with respect to $\partial /\partial q_i'$ on both sides of Equation~(\ref{eq:hamR}), which demonstrates that the Newtonian and Hamiltonian formalisms provide equivalent invariants. For example, writing Equation~(\ref{eq:ham2}) as
\begin{equation}
\label{eq:ham3}
H'({\bf q}',{\bf p}')=R^{2}\hat H({\bf q}',{\bf p}',t)=\bigg[\sum_{i=1}^\nu\frac{1}{2}p_i'^2 +\Phi'({\bf q}')\bigg];
\end{equation}
shows that the Hamiltonian $H'({\bf q}',{\bf p}')$ contains no explicit time dependence and is therefore a dynamical invariant.

To generalize the Newtonian invariant given by Equation~(\ref{eq:inv}) one can simply express the Hamiltonian~(\ref{eq:ham3}) in terms of the original canonical coordinates through the transformation~(\ref{eq:trans_q}) and~(\ref{eq:trans_p}), i.e.
\begin{equation}
\label{eq:ih}
I=\sum_{i=1}^\nu \bigg[ \frac{1}{2}(Rp_i-\dot R q_i)^2 + \frac{1}{2} \ddot R R q_i^2\bigg] + R^2 \Phi({\bf q},t);
\end{equation}
where $R(t)$ is a solution of~(\ref{eq:hamR}), and $[{\bf q}(t),{\bf p}(t)]$ obey the equations~(\ref{eq:h_dotq}) and~(\ref{eq:h_dotp}). For potentials which do not explicitly depend on time (also known as {\it autonomous systems}) $R=1$ is a solution of Equation~(\ref{eq:hamR}), and the total energy becomes an integral of motion.

\subsection{Examples}\label{sec:examples}
The construction of {\it analytical} invariants is only possible in systems where the differential equations that define the coordinate transformation can be de-coupled from the equations of motion. In practice this condition is met when the phase-space coordinates do not appear explicitly in Equation~(\ref{eq:emar_R}) or~(\ref{eq:hamR}).  

Sections~\ref{sec:osc} and~\ref{sec:dirac} discuss a few cases where {\it exact} dynamical invariants can be found using the transformation given by Equations~(\ref{eq:transf}) and~(\ref{eq:trans_h}), while Section~\ref{sec:power} shows that {\it approximate} invariants can be analytically calculated for slowly-varying power-law forces. For more complex systems the construction of analytic invariants will generally require coordinate transformations tailored to the specific scale and/or symmetry of the gravitational field.

\subsubsection{Time-dependent harmonic oscillator}\label{sec:osc}
The case of an harmonic potential which varies with time has a broad range of applications in Quantum Mechanics (see Kaushal 1998 for a review). In Astronomy this potential arises in systems with an homogeneous mass distribution, like in the cores of elliptical galaxies (e.g. Binney \& Tremaine 2008).

For systems with spherical symmetry\footnote{The case of coupled oscillators in two and three dimensions has been studied by Kaushal (1998).} the force in Equation~(\ref{eq:eqmot1}) can be written as $F=-\omega^2(t) r$. Equations~(\ref{eq:eqmot_tind}) and~(\ref{eq:emar_R}) become
\begin{subequations}
\begin{align}
{\ddot r}' +\omega_0^2 r' = 0\label{eq:osc_Q}\\
\ddot R + \omega^2R = \omega_0^2/R^3 \label{eq:osc_R};
\end{align}
\end{subequations}
where $\omega_0$ is a constant. The pair of Equations~(\ref{eq:osc_Q}) and~(\ref{eq:osc_R}) corresponds to an Emarkov (1880) system, wherein the information about the time-evolution of the potential is carried by the auxiliary Equation~(\ref{eq:osc_R}) (see Ray \& Reid 1982). 

From Equation~(\ref{eq:eqmot_tind}) the energy can be written as $I_{HO}=1/2(\d r'/d\tau)^2+1/2 \omega_0^2r'^2$. After substituting Equation~(\ref{eq:osc_R}) into~(\ref{eq:inv}) the energy integral of a time-dependent harmonic oscillator reduces to 
\begin{eqnarray}
I_{HO}=\frac{1}{2}(R\dot r - \dot R r )^2 + \frac{1}{2}\omega_0^2\bigg(\frac{r}{R}\bigg)^2.
\label{eq:IHO}
\end{eqnarray}
where $R(t)$ is a solution of Equation~(\ref{eq:osc_R}).

\subsubsection{Dirac's Cosmology}\label{sec:dirac}
On Dirac's large-number hypothesis (Dirac 1938)\footnote{See Uzan (2003) for a compilation of experimental bounds on the variation of $G$ with time.}
\begin{eqnarray}
\frac{G m_p m_e}{e^2}\simeq 10^{-39}\simeq \frac{e^2}{m_e c^3 t};
\label{eq:dirac}
\end{eqnarray}
where $t$ is the time since the big bang, $m_p$ and $m_e$ are the proton and electron masses, and $e$ is the electron charge. 

In a Universe where the properties of elementary particles remain constant $G=G_0/(H_0 t)$, where $G_0$ and $H_0$ are constants. The force term in Equation~(\ref{eq:eqmot1}) can be simply written as ${\bf F}({\bf r},t)={\bf F}({\bf r})(H_0 t)^{-1}$. 

Lynden-Bell (1982) showed that the choice $R=H_0 t$ in Equation~(\ref{eq:eqmot_trans}) naturally leads to Equation~(\ref{eq:eqmot_tind}). After carrying the inverse transformation of coordinates the energy written as
\begin{eqnarray}
I_D=\frac{1}{2}(H_0t \dot {\bf r} - H_0 {\bf r})^2 + (H_0 t)^2\Phi({\bf r},t);
\label{eq:ID}
\end{eqnarray}
is an exact invariant. Here $\Phi$ is the Newtonian potential in the Dirac's cosmology, i.e. $\Phi({\bf r},t)=-\int \d{\bf r} F({\bf r},t)$ with $G=G_0/(H_0t)$.

\subsubsection{Time-dependent power-law forces}\label{sec:power}
The examples outlined above correspond to time-evolving forces with known constants of motion. Unfortunately, rare are the cases where the trajectory of the particle does not appear explicitly in $R(t)$ (see Lewis \& Leach 1982 and Feix et al. 1987 for further examples related to the harmonic oscillator).

Fortunately, it is relatively simple to construct {\it approximate} invariants that de-couple from the equations of motion for systems that orbit in a slowly-varying gravitational potential. Although the resulting invariants are not exact constants of motion, we shall see below that their evolution is remarkably slow even when forces change on relatively short time scales. As in \S\ref{sec:newton} and \S\ref{sec:hamilton}, the energy integral in the new coordinates turns out to be an invariant of the system.

Let us consider a time-dependent power-law force 
\begin{equation}
\label{eq:plawf}
F(r,t)=-\mu(t)r^n.
\end{equation}
Here the radius expressed in Cartesian coordinates is $r^2=x^2/a^2+y^2/b^2+z^2/c^2$, with $a,b$ and $c$ being constant dimension-less quantities. 

Equations~(\ref{eq:eqmot_trans}) and~(\ref{eq:eqmot_tind}) lead to the auxiliary equation
\begin{equation}
\label{eq:plaw}
\ddot R +\mu R^n r'^{n-1} - \frac{\mu_0}{R^3}r'^{n-1}=0;
\end{equation}
where $\mu_0$ is a constant. 

For forces that change slowly, that is $\epsilon'\equiv \epsilon T_0 \equiv (\dot \mu/\mu_0)T_0\ll 1$, where $T_0$ is the radial period of the orbit at $t=t_0$, it is straightforward to find approximate solutions for Equation~(\ref{eq:plaw}) using a perturbative approach. 
At first order the function 
\begin{equation}
\label{eq:plaw1}
R_1(t)=\bigg[\frac{\mu(t)}{\mu_0}\bigg]^{-1/(n+3)};
\end{equation}
is a solution of Equation~(\ref{eq:plaw}) for $r'\neq 0$. 

By substituting Equation~(\ref{eq:plaw1}) into Equation~(\ref{eq:inv}) the energy in the new coordinate system becomes
\begin{equation}
\label{eq:plawe}
I_n=\frac{1}{2}\bigg(R_1\dot {\bf r} +\frac{\epsilon}{n+3}R_1^{n+4} {\bf r}\bigg)^2 + R_1^2\Phi_1(r,t);
\end{equation}
where $\Phi_1$ can be written as
\begin{equation}
\label{eq:phi1}
\Phi_1(r,t)= 
\begin{cases}
-\frac{\mu}{n+1} r^{n+1} & ,n\ne - 1 \\ 
\mu\ln(r/R_1)& , n=-1 
\end{cases}
\end{equation}

Before we study perturbations at a higher order it is worth examining the solution implied by Equation~(\ref{eq:plaw1}) in more detail. Let us consider the fully analytical case of a time-dependent spherical harmonic oscillator $(n=1$) with frequency $\omega^2(t)=\omega_0^2(1+\epsilon' t/T_0)$, with $\epsilon'=0.01$ and $T_0=2\pi/\kappa_0=\pi/\omega_0=\pi$, where $\kappa_0=2\omega_0$ is the radial frequency of an oscillator. 

\begin{figure}
\begin{center}
 \centerline{\includegraphics[width=84mm]{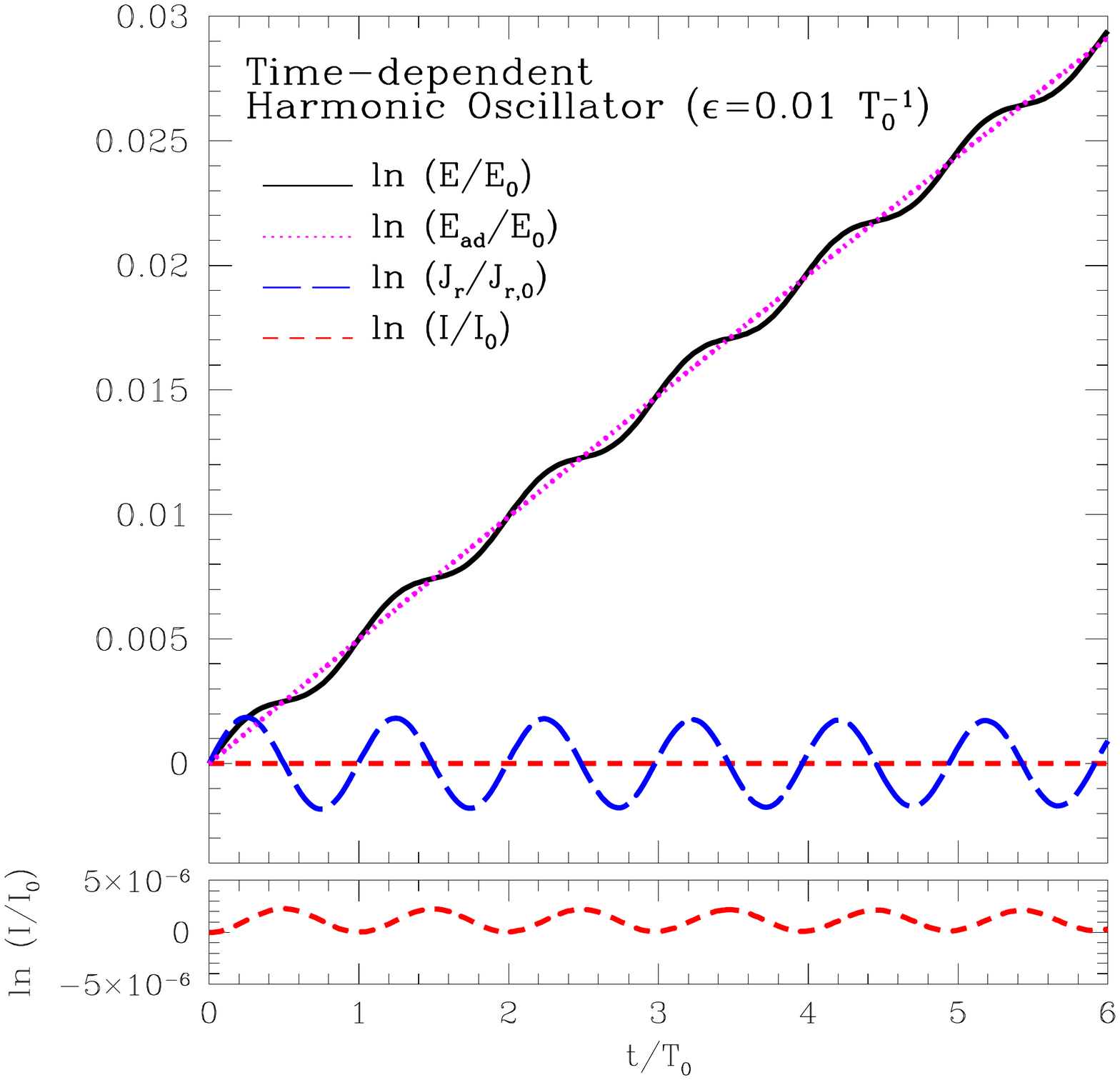}}
\vspace{-2.0truecm}
\caption{Evolution of the orbital energy (black solid line), radial action (blue long-dashed line), and dynamical invariant (red short-dashed line) of a particle moving in a time-dependent harmonic potential $\Phi=\omega^2(t)r^2/2=(1+\epsilon t)r^2/2$, with $\epsilon=0.01/T_0$ and $T_0=\pi/\omega_0=\pi$ being the radial period of the orbit at $t=t_0=0$. The magenta dotted line shows the energy evolution under the adiabatic approximation (Equation~\ref{eq:ead}),  $\ln(E_{\rm ad}/E_0)=2/(3+n)\ln[\Phi(t)/\Phi_0]\approx 1/2\epsilon t$. Note that the variation of the approximate invariant $I$ is of the order $\ln (I/I_0)\sim 10^{-6}$, which is only visible after reducing the scale of the vertical axis (lower panel). }
\label{fig:f1}
 \end{center}
\end{figure}

\begin{figure}
  \centerline{\includegraphics[width=84mm]{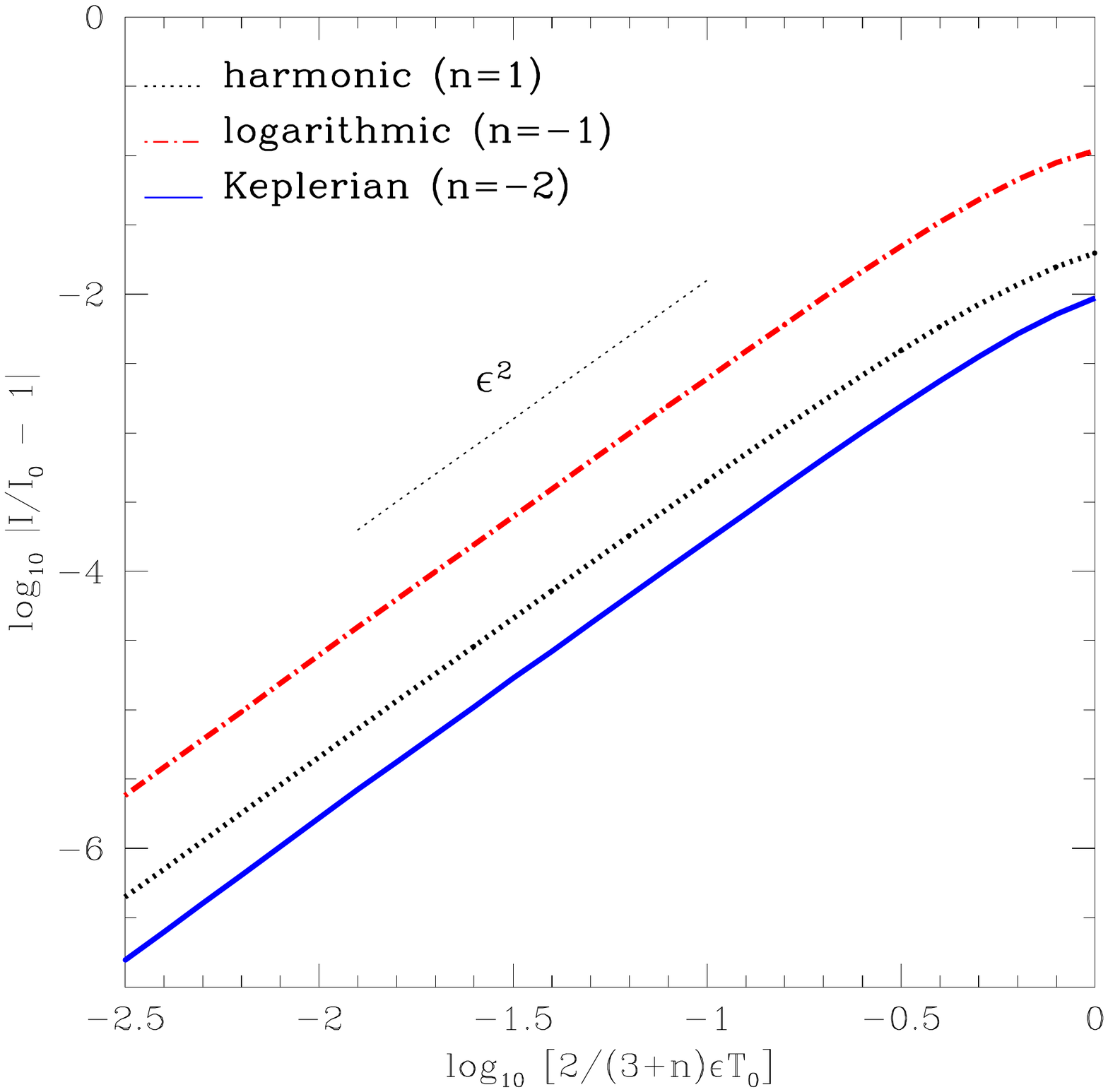}}
\vspace{-2.0truecm}
\caption{Time-averaged evolution of the approximate invariants $I_n$ given by Equations~(\ref{eq:plawe}) and~(\ref{eq:phi1}) as a function of the rate of variation of the gravitational potential, $\epsilon' = (\dot \Phi/\Phi_0) T_0$, where $T_0$ is the radial period of the orbit. We consider orbits in harmonic ($n=1$), logarithmic ($n=-1$) and Keplerian ($n=-2$) spherical potentials with apo- and pericentres $r_{\rm apo}=1$, $r_{\rm peri}=0.4$, respectively. Note that for $\epsilon' \lesssim 1$, the variation of these invariants scales as $|\Delta I_n/I_{n,0}|\sim\mathcal{O}(\epsilon'^2)$.}
\label{fig:eps}
\end{figure}

Fig.~\ref{fig:f1} shows the variation of energy (black solid line) of an orbit with initial energy $E_0=0.58$ and angular momentum $L_0=0.4$, which correspond to orbital apo and pericentres $r_{\rm apo}=1$ and $r_{\rm peri}=0.4$, respectively. In an adiabatic regime the orbital energy of power-law potentials varies as (e.g. Pontzen \& Governato 2012)
\begin{equation}
\label{eq:ead}
\frac{E_{\rm ad}(t)}{E_0}=\bigg[\frac{\mu(t)}{\mu_0}\bigg]^{2/(3+n)};
\end{equation}
which for $n=1$ and $\mu/\mu_0=(1+\epsilon' t/T_0)$ yields $E_{\rm ad}/E_0=(1+\epsilon' t/T_0)^{1/2}\approx 1 + 0.01 t/(2T_0)$ (magenta dotted line). Note that the time-average of the adiabatic energy is $\langle E_{\rm ad}\rangle \approx \langle E \rangle$, showing that the adiabatic approximation is indeed accurate. However, deviations from the adiabatic approximation are clearly visible when the particle is not located at either orbital peri- or apocentre. 
This behaviour can be easily understood using our dynamical invariants. Through Equation~(\ref{eq:inv}) one can express the energy as 
\begin{equation}
\label{eq:ei}
E=\frac{I}{R^2}+ ({\bf r}\cdot\dot{\bf r})\frac{\dot R}{R}-\frac{1}{2}r^2 \bigg(\frac{\ddot R}{R}+\frac{\dot R^2}{R^2}\bigg);
\end{equation} 
From Equations~(\ref{eq:plaw1}) and~(\ref{eq:ead}) the energy evolution calculated at first order is
\begin{eqnarray}
\label{eq:EIHOad}
E\simeq \frac{I}{R_1^2}+({\bf r}\cdot\dot {\bf r})\frac{\dot R_1}{R_1}= I\bigg[\frac{\mu}{\mu_0}\bigg]^{2/(3+n)} -\frac{1}{n+3}\frac{\dot \mu}{\mu}({\bf r}\cdot\dot {\bf r})\\ \nonumber
=E_{\rm ad}-\frac{1}{2}\frac{\dot \omega}{\omega} ({\bf r}\cdot\dot {\bf r}) + \mathcal{O}(\epsilon')^2;
\end{eqnarray}
where the invariant is set to the initial value of the energy, $I=E_0$, and $\mu=\omega^2$. Thus, deviations from the adiabatic approximation in Fig.~\ref{fig:f1} oscillate in phase with the radial motion of the particle and are proportional to $\dot \omega/\omega$.

Similarly, Fig.~\ref{fig:f1} shows that the radial action $J_r$ (dashed lines) oscillates with an amplitude $\sim \mathcal{O}(\epsilon')$. The harmonic potential allows for an analytical expression of the radial action, which can be written as (e.g. Goodman \& Binney 1984) 
\begin{equation}
\label{eq:jr}
J_r =\frac{1}{\pi}\int_{r_{\rm peri}}^{r_{\rm apo}} p_r \d r = \frac{E-\omega L}{2\omega};
\end{equation}
where $L$ is the angular momentum of the particle. Inserting Equation~(\ref{eq:EIHOad}) in~(\ref{eq:jr}) yields 
\begin{equation}
\label{eq:jrosc}
J_r =J_{r,0} -\frac{1}{4}\frac{\dot \omega}{\omega^2}({\bf r}\cdot\dot {\bf r}) + \mathcal{O}(\epsilon')^2.
\end{equation}
The first term $J_{r,0}=(E_0-\omega_0 L)/(2\omega_0)$ is a constant, while the right-hand term oscillates in phase with the radial motion of the orbit. Notice, however, that the term accompanying $({\bf r}\cdot\dot {\bf r})$ vanishes when averaged over a full orbital revolution. Therefore, the time-average $\langle J_{r}\rangle=1/t\int_0^{t} \d t' J_{r}(t')$ is conserved at order $\mathcal{O}(\epsilon'^2)$ for $t\gg T_0=2\pi/\omega_0$.

In contrast the evolution of the approximate invariants $I_n$ is at order $\mathcal{O}(\epsilon'^2)$ {\it along the phase-space path of the particle motion}. This is a remarkable result given that $R_1$ is a solution of Equation~(\ref{eq:plaw}) at order $\mathcal{O}(\epsilon')$. To understand the higher-order behaviour of $R(t)$ let us construct a function $R\approx R_1+\delta R_2$, where $R_1$ is a solution of Equation~(\ref{eq:plaw1}) and $\delta\ll 1$, and $R_2$ a residual function. Inserting $R$ in Equation~(\ref{eq:plaw}), isolating the terms proportional to $\delta$, and neglecting those at higher order yields the second-order differential equation
\begin{equation}
\label{eq:plaw2}
\ddot R_2 +\mu_0(n+3)r'^{n-1} R_2=0.
\end{equation}
Note that for $n=1$ this is the equation of a harmonic oscillator with frequency $2\sqrt{\mu_0}$, which also corresponds exactly to the radial frequency of the time-independent potential associated with Equation~(\ref{eq:plawf}), i.e. $\kappa=2\omega_0=2 \sqrt{\mu_0}$.
Indeed, a similar result is obtained for the Keplerian case ($n=-2$) if we approach $r'\approx a$, where $a$ is the semi-major axis of the orbit. In this case we find that $R_2$ follows a cycle with a frequency $\sqrt{\mu_0/ a^{3}}$, or a period $T=2\pi\sqrt{a^3/\mu_0}$, which corresponds to the radial period of a Keplerian orbit. 

In general, Equation~(\ref{eq:plaw2}) shows that $R_2$ oscillates approximately in phase with the radial motion of the particle about the centre of the power-law force field. Hence, if the time dependence of the force evolves slowly, the averaged contribution of the terms accompanying $\ddot R$ in Equation~(\ref{eq:inv}) can be safely neglected.

As a result we find that in slowly-varying potentials ($\epsilon'\lesssim 1$) the approximate invariants given by Equations~(\ref{eq:plawe}) and~(\ref{eq:phi1}) are accurate at order $\mathcal{O}(\epsilon'^2)$, even if the scale factor $R_1$ is a solution of Equation~(\ref{eq:plaw}) at order $\mathcal{O}(\epsilon')$. This result is illustrated in Fig.~\ref{fig:eps}, where the mean variation of $|\Delta I/I_0|$ is plotted against the rate of potential change $2/(3+n) \epsilon'$. The factor $2/(3+n)$ ensures that the fractional change of energy is approximately the same for all the orbits considered here. Note that on average $I$ varies by a small amount, $|\Delta I /I_0|\lesssim 0.1$, even in cases where the adiabatic approximation does {\it not} hold and the potential evolves on a time-scale comparable to the radial period of the orbit, i.e. $\epsilon T_0\sim 1$.

\section{An Application: Accreted substructures in time-dependent galactic potentials }\label{sec:appl}
Hierarchical theories of structure formation propose that galaxies form through the accretion of smaller, gravitationally-bound bodies (White \& Rees 1978). A natural prediction from this scenario is the presence of dynamical fossils in the present-day configuration space of the Milky Way. 

The integral-of-motion space may offer the best chances to uncover the hierarchical build-up of our Galaxy. Here accreted stars are expected to distribute in tight clumps rather than homogeneously, reflecting the fact that they were originally bound to low-mass systems which did not form in situ. However, this approach has a strong limitation: as the Galaxy grows hierarchically, so does its overall gravitational potential. Under such circumstances none of the integrals of motion is conserved. Hence, a natural question arises as to how tidal substructures evolve in the integral-of-motion space under a time-dependent potential. Below we attempt to tackle this issue using the dynamical invariants constructed in the previous Sections. 

\subsection{Orbital diffusion}\label{sec:scatt}
Before we attempt to answer this question in detail it is worthwhile to study simple models that share the essential features of dynamical fossils in a time-dependent potential. For example, let us consider two particles moving in a spherical potential $\Phi(r,t)=\mu(t)h(r)$ which varies slowly with time. These particles do not interact gravitationally, so their motion is entirely governed by $\Phi$. We construct our experiment so that at time $t=t_0$ both particles move on the same orbit but are located at different radii. Hence $\Delta E(t_0)=E_1(t_0)-E_2(t_0)=0$, that is $v_1^2/2 + \mu(t_0)h(r_1)= v_2^2/2 + \mu(t_0)h(r_2)$. We now integrate their orbits forward in time until both particles exchange their radial location. At that particular time, say $t=t_1$, their energies are $E_1(t_1)\approx v_2^2/2 + \mu(t_1)h(r_2)$ and $E_2(t_1)\approx v_1^2/2 + \mu(t_1)h(r_1)$, where it is assumed that the potential varies so slowly as to leave the orbital velocity at both radii unchanged. It is straightforward to show that at $t=t_1$ the energies of the two particles differ by the amount $\Delta E\approx [\mu(t_1)-\mu(t_0)][h(r_2)-h(r_1)]$.

This simple model illustrates two interesting features of tidal substructures evolving in time-dependent potentials that will become more obvious below. First, notice that $\Delta E$ depends on the relative position of the particles. For tidal debris composed of many particles distributed over a large phase-space volume the growth of the host potential must necessarily lead to a progressive {\it diffusion} of orbital energies\footnote{A similar mechanism operates in the action-angle space, see Pontzen \& Governato (2013).}. Over time this process may smooth out (and perhaps even efface) pre-existing clumps in the integral-of-motion space. Also, the fact that the length of tidal tails oscillates between peri- and apocentre, namely stretching over a large range of galactocentric distances as the progenitor moves close to orbital perincentre and piling-up at orbital apocentre (see Dehnen et al. 2004 for a beautiful illustration of this cycle), suggests that the diffusion process, rather than being gradual, may follow a cyclic evolution in phase with the radial motion of the progenitor system\footnote{Orbital diffusion must not be confused with phase-space mixing. The former tends to increase the phase-space volume available to a particle ensemble, while the latter is the process through which this volume is filled.} . Sections~\ref{sec:nbody} and~\ref{sec:thermo} show that dynamical invariants provide a useful tool to understand the intricacies of this complex process.


\subsection{Entropy evolution}\label{sec:entroev}
The simplistic model of \S\ref{sec:scatt} assumes that all particles moving on a tidal stream follow the same orbit. In reality, tidal streams have an involved orbital structure (e.g. K\"upper et al. 2010; Eyre \& Binney 2011) whose complexity increases in proportion to the mass of the progenitor system (Pe\~narrubia et al. 2006; Choi et al. 2007). 

Let us define $\pi(E,{\bf r},\dot {\bf r},t)$ as the probability to find a star with an energy $E$ at a phase-space location $({\bf r},\dot {\bf r})$ at the time $t$. The probability function is normalized so that $\int \int \int \pi(E,{\bf r},\dot {\bf r},t)\d E \d^3 r \d^3 v =1$.

To measure the orbital scatter introduced by the time dependence of the host potential it is useful to define the entropy associated with the energy distribution as 
\begin{equation}
\label{eq:entro}
\mathcal{H}_E(t)\equiv -\int\int\int \pi(E,{\bf r},\dot {\bf r},t)\ln \pi(E,{\bf r},\dot {\bf r},t)\d E  \d^3 r \d^3 v .
\end{equation}
Stars that are strongly clumped in the integral-of-motion space will have low values of $\mathcal{H}_E$, whereas the entropy of stars moving on loosely correlated orbits will be comparable to that of the smooth stellar background.

The time-evolution of the entropy can be calculated through the energy invariants constructed in \S\ref{sec:newton} and \S\ref{sec:hamilton}. Using Equation~(\ref{eq:ei}) and expanding the energy distribution at order $\mathcal{O}(\epsilon')$ the probability distribution becomes
\begin{equation}
\label{eq:pfun}
\pi (E,{\bf r},\dot {\bf r},t)\simeq \pi (R^{-2}I,{\bf r},\dot {\bf r},t) + ({\bf r}\cdot\dot{\bf r})\frac{\dot R}{R}\pi'(R^{-2}I,{\bf r},\dot {\bf r},t).
\end{equation} 
where the prime denotes derivative with respect to energy, i.e. $\pi'=\d \pi/\d E$, evaluated at $E=R^{-2}I$. Neglecting the terms $\mathcal{O}(\epsilon'^2)$ and after some algebra the entropy defined in Equation~(\ref{eq:entro}) becomes
\begin{eqnarray}
\label{eq:entro1}
\mathcal{H}_E(t)\simeq -\int [\pi(I)\ln \pi(I) + \pi(I)\ln R^2]\d I \\ \nonumber
-\frac{\dot R}{R}\int\int\int ({\bf r}\cdot\dot{\bf r})\pi'(R^{-2}I,{\bf r},\dot {\bf r},t) [1+\ln \pi(R^{-2}I,{\bf r},\dot {\bf r},t)]\d I \d^3 r \d^3 v \equiv \\ \nonumber
\mathcal{H}_I - 2\ln R(t) + \mathcal{H}_{\rm osc}(t);
\end{eqnarray} 
where $\mathcal{H}_I$ is the entropy associated with the invariant energy distribution and has, therefore, a constant value. The right-hand term of Equation~(\ref{eq:entro1}), $\mathcal{H}_{\rm osc}$, depends on the distribution of stars in phase space and has in general no analytical expression. However, its time evolution is well defined. 
\begin{figure*}[!]
\includegraphics[width=160mm, height=240mm ]{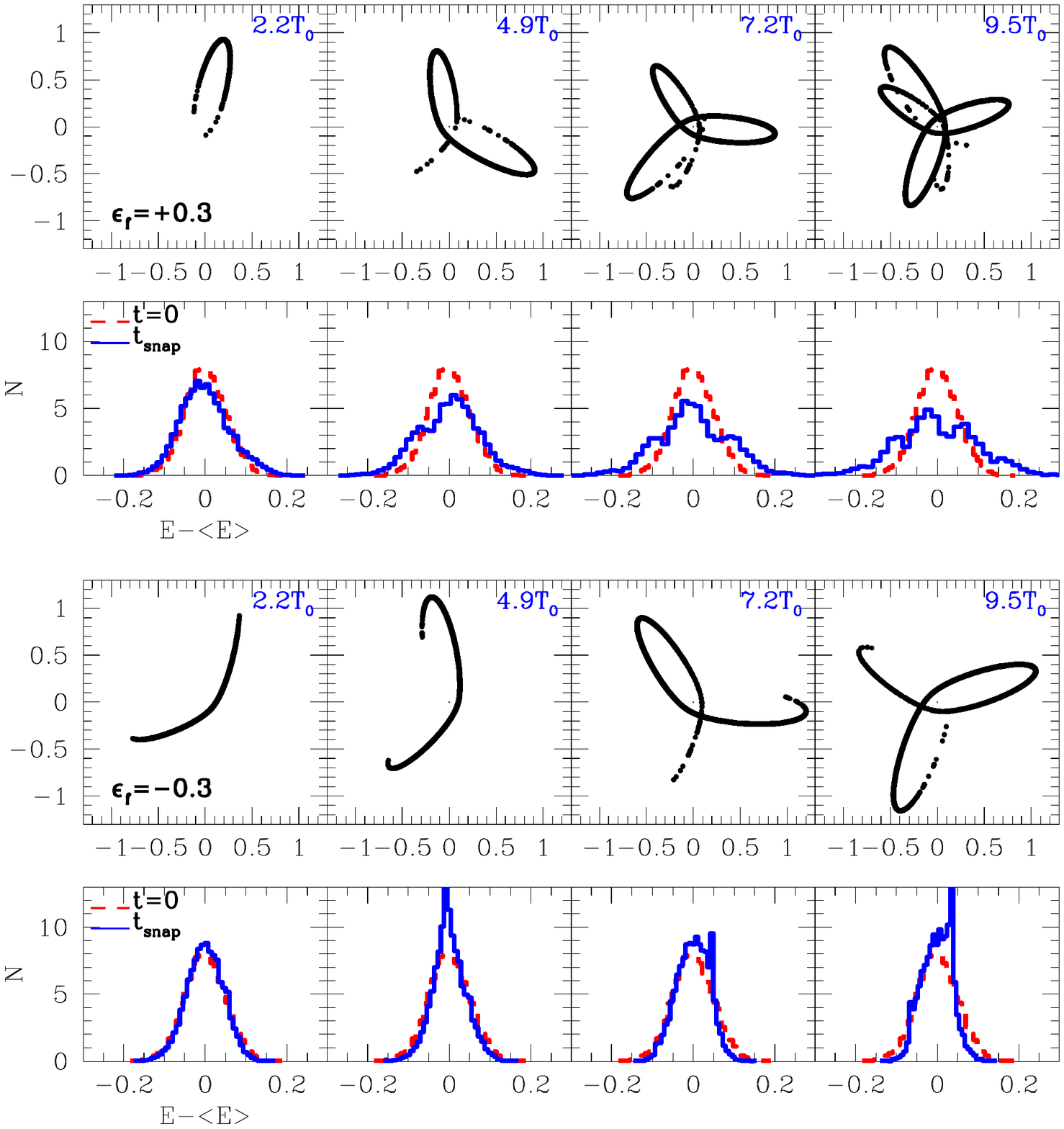}
\vspace{-2.0truecm}
\caption{{\it First and third rows from the top}: Projection of stream particles onto the orbital plane at different snapshots. The stream model is composed of $N_\star=10^4$ particles orbiting in the potential of Equation~(\ref{eq:potlog}) with $\epsilon_f=+0.3$ and -0.3. The integration time is $t_f=10T_0$, where $T_0$ is the radial period of an orbit with apocentre $r_{\rm apo}=1$ and $r_{\rm peri}=0.09$ at $t=0$. Note that in decreasing potentials $(\epsilon_f<0)$ unbound particles spread out on the orbital path on longer time-scales. {\it Second and fourth rows from the top}: Distribution of orbital energies at the corresponding snapshots. Initially, the distribution is Gaussian with a dispersion $\sigma_s=0.05$. Note that in time-dependent potentials the energy distribution of tidal streams tends to thicken/narrow depending on whether the potential grows/shrinks with time.}
\label{fig:stream1} 
\end{figure*}

Let us illustrate the evolution of $\mathcal{H}_{\rm osc}$ by considering stream particles with an energy distribution that is initially separable in space, an assumption which in less inaccurate for tidal debris from low-mass progenitors evolving in static potentials (see Pe\~narrubia et al. 2012). Under this approximation all particles on a given energy surface $I$ have equal probability to move with a radial velocity $\dot{\bf r}\cdot {\bf e}_r$, where ${\bf e}_r$ is the radial unit vector. It is then straightforward to show that the integral $\int \pi'(1+\ln \pi)\d I=0$, and thus $\mathcal{H}_{\rm osc}=0$. 
However, as stream particles spread out on their orbital paths a correlation between the radial velocity and the relative location in phase-space will arise as a result of the orbital diffusion process discussed in Section~\ref{sec:scatt}. During the early phases of the stream evolution most particles move on orbits that remain close to that of the progenitor. Thus, at early times the term $\mathcal{H}_{\rm osc}$ fluctuates in phase with the radial motion of the progenitor system. 
If the host potential evolves adiabatically the orbital periods of the stellar orbits are much shorter than the time-scale of the potential evolution, i.e. $T_0\ll (\dot \Phi/\Phi)^{-1}=\epsilon^{-1}$. Under the assumption that $\dot R/R$ remains approximately constant during a full orbital revolution, the oscillations of $\mathcal{H}_{\rm osc}$ are cyclic and the time-averaged evolution of the entropy becomes
\begin{eqnarray}
\label{eq:entroav}
\langle \mathcal{H}_E \rangle(t)= \frac{1}{t}\int_0^{t}\d t \mathcal{H}_E(t)= \mathcal{H}_I - 2\ln R(t) + \mathcal{O}(\epsilon'^2);
\end{eqnarray}
where $t\gg T_0$.

As tidal streams progressively fill the phase-space volume available to their orbits the number of particles moving toward apocentre (${\bf r }\cdot\dot{\bf r}>0$) tends to approach that moving away from it (${\bf r }\cdot\dot{\bf r}<0$). Hence, the probability of finding a particle in the energy interval $(E, E+\d E)$ at a given time $t$ becomes independent of the sign of $({\bf r}\cdot\dot{\bf r})$. By symmetry the right-hand term of Equation~(\ref{eq:entro1}) tends to $\lim_{\sum ({\bf r}\cdot\dot {\bf r})\rightarrow 0}\mathcal{H}_{\rm osc}=0$, and through comparison with Equation~(\ref{eq:entroav}) we find that the entropy of the microcanonical ensemble tends toward its time-averaged value as substructures mix in phase space. Henceforth we shall consider that tidal debris have reached a state of {\it dynamical equilibrium} if $\mathcal{H}_E-\langle \mathcal{H}_E\rangle=0$. Note that in this limit the entropy evolves as $\lim_{t\rightarrow \infty}\Delta \mathcal{H}_E=\langle \Delta \mathcal{H}_E\rangle\approx -2\ln R$, independently of the initial particle distribution in the integral of motion space. 

Perhaps the most remarkable result inferred from Equation~(\ref{eq:entroav}) is the increasing entropy of substructures orbiting in a growing potential $(R<1$). In cosmologically-motivated galaxy models, which have a triaxial shape and only admit one integral of motion (the orbital energy), this result suggests that the dynamical signatures of accretion may be erased as a result of the hierarchical growth of the host galaxy. The inescapable consequence of the dynamical deposition, and progressive dissolution, of tidal clumps in the integral-of-motion space is the formation of a {\it smooth} stellar halo. A completely smooth galaxy, however, will never emerge from this process because the same mechanism that removes tidal substructures, i.e. the potential growth through the merger of smaller bodies, is also responsible for the formation of new ones.
 
In galaxies with shrinking potentials ($R>1$) the evolution of entropy follows the reverse trend. Here tidal clumps tend to become more prominent with time. Examples of systems which have a decreasing potential may be found, for example, in globular clusters or satellite galaxies losing mass to tides\footnote{However, the presence of tidal clumps in these systems has a transient nature. Eventually the tidal stripping of stellar material effaces all pre-existing substructures (e.g. Pe\~narrubia et al. 2009; Sales et al. 2010).}.

\subsection{$N$-body models in a logarithmic potential}\label{sec:nbody}
It is worth illustrating the above results by running restricted $N$-body models of unbound substructures evolving in a time-dependent logarithmic potential. Let us construct an idealized tidal stream model composed of $N_\star$ particles which do not interact gravitationally among themselves and follow a Gaussian energy distribution, i.e.
\begin{equation}
\label{eq:gauss}
\pi(E,t_0)=\frac{1}{\sqrt{2\pi \sigma_s^2(t_0)}}\exp\{-[E-E_{s}(t_0)]^2/[2\sigma_s^2(t_0)]\};
\end{equation}
where $E_s(t_0)=v_s^2/2+\Phi({\bf r}_s,t_0)$ is the mean orbital energy and $\sigma_s(t_0)$ the energy dispersion at $t=t_0$. Following K\"upper et al. (2012) all particles are placed initially at orbital apocentre with a common velocity vector, ${\bf v}_s=(v_r,v_t)$, where the radial component is $v_r=\hat{\bf r}\cdot{\bf v}_s=0$, and the tangential component is $v_t=\sqrt{2[E_s-\Phi({\bf r}_s,t_0)]}$. Note that the eccentricity of the orbits is set by our choice of $v_t$. With this set-up it is straightforward to generate a sample of apocentres (e.g. using a rejection method) so that the initial energy distribution follows Equation~(\ref{eq:gauss}).

The host galaxy is modelled as an isothermal sphere whose potential evolves linearly with time, i.e.
\begin{equation}
\label{eq:potlog}
\Phi(r,t)=\mu(t)\ln r\equiv \bigg(1+\epsilon_f \frac{t-t_f}{t_f}\bigg)\ln r.
\end{equation}
Hence, setting $t_0=0$, this potential varies from $\mu(0)=1-\epsilon_f$, to $\mu(t_f)=1$ within an integration time $t_f=10 T_0$, where $T_0$ is the radial period of an orbit with energy $E_s(0)$ and angular momentum $L_s=r v_t=v_t$. Note that $\epsilon_f$ can be either positive or negative.

The energy invariant can now be easily calculated by choosing $n=-1$ in Equation~(\ref{eq:plawe}) and~(\ref{eq:phi1}), and computing the scale factor from Equations~(\ref{eq:plaw1}) and~(\ref{eq:potlog}) as
 \begin{equation}
\label{eq:Rlog}
R(t)= \bigg[\frac{t_f+\epsilon_f (t-t_f)}{t_f(1-\epsilon_f)}\bigg]^{-1/2}.
\end{equation}

Fig.~\ref{fig:stream1} shows snap-shots of the time-evolution of a tidal stream with an energy dispersion $\sigma_s(0)=0.05$ orbiting on an eccentric orbit ($v_t=0.24 v_c[0]$, where $v_c[0]=\sqrt{\mu[0]}$ is the circular velocity of the host at $t=0$) in a potential that varies at a rate $\epsilon_f=+0.3$ (two upper-most panels) and $\epsilon_f=-0.3$ (two lower-most panels). The projection of the particles onto the orbital plane shows that particles progressively spread out on the orbital path of the ``progenitor'' system, which follows an orbit with energy $E_s$ and angular momentum $L_s$, leading to the formation of tail-like structures. 
Because the dynamical time scales as $t_{\rm dyn}\propto \mu^{-1/2}$ the formation of tails accelerates in potentials that grow with time, and slows down in potentials that shrink with time. 
The second row of panels show the evolution of the energy distribution of a tidal substructure orbiting in a growing potential. 
These models illustrate the `dissolution' of tidal substructures through the orbital diffusion process outlined in \S\ref{sec:scatt}. The bottom row shows that in shrinking potentials the diffusion process appears to {\it reverse}, i.e. the energy distribution of tidal debris tends to become narrower with time. 
\begin{figure}
\centerline{\includegraphics[width=84mm]{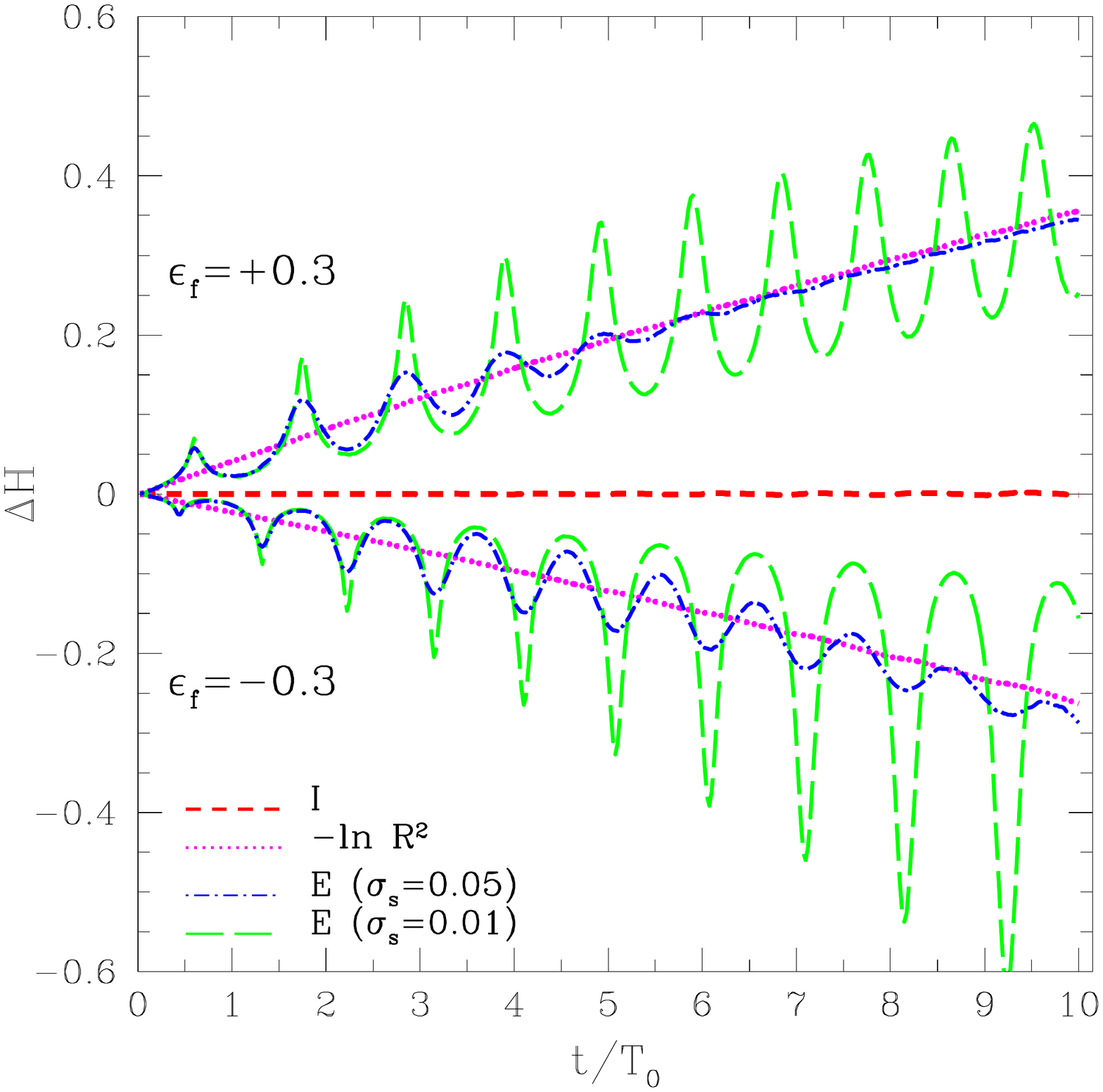}}
\vspace{-2.0truecm}
\caption{Entropy variation of tidal streams orbiting in a logarithmic potential that evolves at a constant rate $\epsilon_f=(\dot \Phi/\Phi) t_f$, where $t_f=10 T_0$ is the orbital integration time (see Fig.~\ref{fig:stream1}). As expected, the entropy associated to the energy invariant ($\mathcal{H}_I$, red dashed line) remains close-to constant throughout the simulation. In contrast, the entropy associated to the energy distribution of tidal streams ($\mathcal{H}_E$) either increases or decreases depending on whether the host potential grows ($\epsilon_f>0$) or shrinks ($\epsilon_f<0$) with time. As expected from Equation~(\ref{eq:entroav}), its averaged evolution (magenta dotted lines) is $\langle \mathcal{\Delta H}_E\rangle \approx -2\ln R(t)$, where $R(t)$ is given by Equation~(\ref{eq:potlog}). The entropy of tidal streams undergoes periodic oscillations in phase with the radial period of the progenitor's orbit. Comparison between the models with $\sigma=0.01$ (green long-dashed lines) vs. those with $\sigma=0.05$ (blue dotted-dashed lines) shows that the amplitude of the oscillations increases for streams that have initially a low energy dispersion.}
\label{fig:entrotime}
\end{figure}

The evolution of the entropy associated to the above models 
is shown in Fig.~\ref{fig:entrotime} (blue dotted-dashed lines). As expected from Equation~(\ref{eq:entroav}), the entropy of tidal debris oscillates about a time-average value $\langle \Delta \mathcal{H}_E\rangle=-2\ln R$, where $R$ is given by Equation~(\ref{eq:Rlog}) (magenta dotted lines). The amplitude of the oscillations decreases as the stream particles approach dynamical equilibrium. However, the damping process is considerably slower for substructures that are dynamically `cold' ($\sigma_s=0.01$, green long-dashed lines), or if the tidal stream particles orbit in a decreasing potential ($\epsilon_f<0$). This suggests that not all substructures may reach dynamical equilibrium within a Hubble time. 
It is also worth noting that the entropy associated to the energy invariants (red dashed line) remains remarkably constant throughout the evolution of these models. The accuracy of our energy invariant, $I$, can be estimated directly from Fig.~\ref{fig:eps}. Measuring the fractional variation of the logarithmic potential as $\epsilon= \epsilon_f/t_f=0.03 T_0^{-1}$, we find $|\Delta I/I_0|\lesssim 10^{-4}$ and $|\Delta \mathcal{H}_I|\lesssim 10^{-3}$.

\subsection{Thermodynamics of tidal substructures}\label{sec:thermo}
Statistical mechanics provide an alternative physical description of the macroscopic properties of gravitating systems in dynamical equilibrium. In classical thermodynamics the probability of finding a particle in the energy interval $(E, E+\d E)$  at a given time $t$ can be calculated as $\pi_{\rm th}(E,t)=g(E,t)f(E,t)$, where $g(E,t)$ is the volume of phase space of the constant energy surface $E=H$; and $f(E,t)$ is the distribution function. For simplicity let us again adopt the potential~(\ref{eq:potlog}), which corresponds to a self-gravitating isothermal sphere in dynamical equilibrium. In this potential both functions $g(E,t)$ and $f(E,t)$ have analytical expressions. 

The density of states is
\begin{eqnarray}
\label{eq:gammae}
g(E,t)=\int \d^3 r \d^3 v \delta [E-H(t)] \\ \nonumber
=(4\pi)^2 \int_0^{r_m(E,t)} r^2 \sqrt{2[E-\Phi(r,t)]}\d r=A\sqrt{\mu(t)}\exp[3 E / \mu(t)];
\end{eqnarray}
where $r_m(E,t)$ denotes the radius at which $E=\Phi$ at time $t$ (e.g. Binney \& Tremaine 2008); and $A=8/9 \pi^2\sqrt{6\pi}$. 

Using the classical definition of thermal entropy, $S=\ln g(E,t)$, the temperature of the sphere can be calculated as
\begin{equation}
\label{eq:temp}
\mathcal{T}=\bigg(\frac{\d S}{\d E}\bigg)^{-1}=\frac{\mu(t)}{3}.
\end{equation}
If the system is composed of particles with mass $m$ with mean kinetic energy $\langle 1/2m v^2\rangle$, the temperature is typically measured as $3/2 K_B \mathcal{T}= \langle 1/2 m v^2\rangle $, where $K_B$ is Boltzmann's constant (e.g. Feynman 1963). Therefore, by defining the (one-dimensional) velocity dispersion of an isothermal sphere as $\sigma^2=\langle v^2\rangle/3$, Equation~(\ref{eq:temp}) becomes $\mu=3 \mathcal{T}= (3m/K_B) \sigma^2$. 

The phase-space distribution corresponding to the potential~(\ref{eq:potlog}) is (Binney \& Tremaine 2008)
\begin{equation}
\label{eq:fe}
f(E,t)=\frac{1}{[\pi\mu(t)]^{3/2}}\exp[-2 E/\mu(t)].
\end{equation}
Thus, from Equations~(\ref{eq:gammae}) and ~(\ref{eq:fe}) the probability $\pi_{\rm th}(E,t)$ can be written as
\begin{equation}
\label{eq:pth1}
\pi_{\rm th}(E,t)=g(E,t)f(E,t)=\frac{B}{\mu(t)}\exp[E/\mu(t)];
\end{equation}
with $B$ chosen so that the normalization of the probability function is $\int \pi_{\rm th}(E,t)\d E=1$. 

Let us now compare the entropy derived using the standard methods of equilibrium statistical mechanics and that resulting from the construction of dynamical invariants. Substituting Equation~(\ref{eq:pth1}) into~(\ref{eq:entro}) and changing the integration variable to $I=(\mu_0/\mu)E$, so that $\pi_{\rm th}(E)\d E=\pi_{\rm th}(I)(\mu_0/\mu) \d I$, we find that the entropy associated to $\pi_{\rm th}$ evolves as
\begin{equation}
\label{eq:hth}
\mathcal{H}_{E,{\rm th}}(t)=-\int \d E \pi_{\rm th}(E,t) \ln \pi_{\rm th}(E,t)=\mathcal{H}_{I,{\rm th}} +\ln [\mu(t)/\mu_0].
\end{equation}
It is straightforward to show that the time-averaged entropy of tidal streams corresponds to the thermodynamical entropy of the host. For logarithmic potentials, $n=-1$, Equation~(\ref{eq:plaw1}) becomes $R=(\mu/\mu_0)^{-1/2}$. Comparison of Equations~(\ref{eq:hth}) and~(\ref{eq:entroav}) shows that $\Delta \mathcal{H}_{E,{\rm th}}=\langle \Delta \mathcal{H}_E\rangle$. 
Hence, both descriptions of entropy become identical in the limit of dynamical equilibrium, i.e. when the number of particles on an energy surface $E=H$ moving outwards is equal to that moving inwards, $\lim_{\sum ({\bf r}\cdot\dot {\bf r})\rightarrow 0}\mathcal{H}_{E}=\mathcal{H}_{E,{\rm th}}$.

Fig.~\ref{fig:entrotime} can now be re-interpreted in terms of thermodynamical temperatures. Comparison of Equation~(\ref{eq:entroav}) and~(\ref{eq:temp}) shows that the temperature of a logarithmic potential evolves as
\begin{equation}
\label{eq:avert}
\bigg\langle \frac{\mathcal{T}}{\mathcal{T}_0} \bigg\rangle=\exp[\langle \Delta \mathcal{H}_E \rangle ];
\end{equation}
 where brackets denote average over time. In a hierarchical galaxy formation framework this implies that galaxies heat up as they build up mass through the accretion of smaller bodies. 
In contrast, the entropy, and thus the temperature, of tidally-stripped objects drops progressively as they lose mass to tides\footnote{Note that the second Law of thermodynamics mandates that the total entropy of the host-satellite system must increase. This happens through the formation of tidal tails.}.
 Thermodynamically these systems behave as if they were in contact with hot and cold thermal baths, respectively\footnote{The use of `hot' and `cold' may sound counter-intuitive here. By definition satellites are colder than the parent galaxy, yet it appears as if they were in contact with a cold bath. Similarly, although the Universe is colder than the host galaxy, it acts as a hot bath. The source of confusion, as usual, can be traced back to the negative heat capacity of gravitationally-bound systems (e.g. Padmanabhan 1990).}.

Although the thermal bath analogy is helpful, it fails to provide a correct description of the dynamical evolution of tidal substructures that have not yet spread out on their orbital paths and are, therefore, out of dynamical equilibrium. Indeed, Fig.~\ref{fig:entrotime} shows that, far from changing monotonically as one would expect for systems in contact with a thermal bath, the temperature of tidal streams fluctuates about that of the host galaxy. In terms of statistical mechanics this implies that the amount of energy required to change the temperature (i.e. the heat capacity) varies along the orbital path. 

This odd property can be easily understood through the energy invariants. Defining the stream energy as $E_s=\int\int\int \d E \d^3 r\d^3 v \pi(E,{\bf r},\dot {\bf r},t) E$ and using Equation~(\ref{eq:ei}) it is straightforward to show that at the early stages of the stream evolution, i.e. when the phase-space distribution of stream particles remain close to the phase-space location of the progenitor system, this quantity becomes
\begin{equation}
\label{eq:Es}
E_s\simeq I_s \frac{\mathcal{T}}{\mathcal{T}_0} - \frac{1}{2}({\bf r}_s\cdot\dot {\bf r}_s)\bigg(\frac{\mathcal{\dot T}}{\mathcal{T}}\bigg);
\end{equation}
where the scale factor $R$ has been expressed in terms of the thermodynamical definition of temperature given by Equation~(\ref{eq:temp}), i.e. $R=(\mu/\mu_0)^{-1/2}=(\mathcal{T}/\mathcal{T}_0)^{-1/2}$; and $I_s$ is the invariant energy of the stream. From Equation~(\ref{eq:Es}) the first-order variation of the heat capacity is 
\begin{equation}
\label{eq:cv}
C=\frac{\d E_s}{\d \mathcal{T}}\simeq \frac{I_s}{\mathcal{T}_0}+ \frac{1}{2}({\bf r}_s\cdot\dot {\bf r}_s)\bigg(\frac{\mathcal{\dot T}}{\mathcal{T}^2}\bigg);
\end{equation}
For tidal substructures that are energetically bound ($I_s<0$) the heat capacity has a negative sign. Throughout the orbit of tidal substructures the heat capacity oscillates about a constant value $\langle C\rangle =I_s/\mathcal{T}_0$. In growing potentials ($\mathcal{\dot T}/\mathcal{T}_0>0$) the quantity $C-\langle C\rangle$ is negative toward pericentre and positive toward apocentre. In shrinking potentials ($\mathcal{\dot T}/\mathcal{T}_0<0$) the cycle reverses. Note also that the right-hand term of Equation~(\ref{eq:cv}) is proportional to $\mathcal{\dot T}/\mathcal{T}^2$. Therefore, the fluctuations in temperature of tidal streams are bound to damp out as the temperature of the host rises. In contrast, if the temperature of the host drops the right-hand term of Equation~(\ref{eq:cv}) grows with time and the convergence toward dynamical equilibrium cannot be guaranteed (see Fig.~\ref{fig:entrotime}). 

The reason why the thermodynamical definition of entropy does not reproduce this peculiar behaviour can be traced back to the definition of density states itself. Equation~(\ref{eq:gammae}) presumes that the particles of an ensemble distribute throughout volumes of constant energy surfaces $E=H$. Maximization of entropy $S$ is thus equivalent to the maximization of the phase volume available to those particles. However, Fig.~\ref{fig:stream1} shows that tidal substructures violate this assumption. Indeed, the phase-space volume filled by tidally-stripped particles fluctuates with time and, in general, it takes several orbital revolutions until the particles spread out on the available phase-space volume of the orbit. 
Not surprisingly we find that the temperature of tidal streams approaches the thermodynamical value in the limit of dynamical equilibrium, that is 
\begin{equation}
\label{eq:tempentro}
\lim_{\sum_i ({\bf r}\cdot\dot {\bf r})\rightarrow 0}\bigg(\frac{\mathcal{T}}{\mathcal{T}_0}\bigg)=\frac{1}{R^{2}}.
\end{equation}
Note, however, that for low-mass progenitors this limit is reached on time scales longer than the age of the Universe!

\subsection{Smooth vs. clumpy stellar halo}\label{sec:detect}
The previous Sections have laid out the evolution of dynamical fossils in a time-dependent potential. The fate of stellar substructures that form through the merger of small bodies is to be effaced by the hierarchical growth of the host potential. Hence, in a hierarchical galaxy formation framework the same dynamical mechanism that leads to the proliferation of tidal substructures, i.e. the accretion of gravitationally-bound systems, is also responsible for their progressive removal. The `smooth' galactic component arises as an inescapable by-product of this cycle.

Given that all substructures are on average equally affected by orbital diffusion (see \S\ref{sec:entroev}), whether or not dynamical fossils can be detected in the present-day configuration space will mainly depend on three (typically unknown) factors, namely, the time-dependence of the Milky Way potential, the `age' of tidal substructures (defined as the look-back time since these stars were tidally stripped from the progenitor system), and the initial distribution of tidal debris in the integral-of-motion space. A quantitative description of the evolution of tidal substructures appears, therefore, an impossibly difficult task. However, it is feasible to construct simple toy models that share the essential features of these systems and hence offer useful insight into the problem at hand.

Let us begin by adopting a cosmologically-motivated mass growth for our host galaxy. In numerical (collisionless) simulations of structure formation the average mass evolution of galactic haloes follows a relatively simple function, 
\begin{equation}
\label{eq:mcosmo}
 \mu(z)=\mu_0\exp[-2 z/(1+z_c)];
\end{equation}
where $z_c=c_0/4.1-1$ is the formation redshift and $c_0$ is the virial concentration at $z=0$ (Wechsler et al. 2002). Adopting a fiducial Milky Way mass of $\mu_0=10^{12}M_\odot$ and using the mass-concentration relationship observed in cosmological simulations (Macci\`o et al. 2007) yields $z_c\simeq 1.44$. Hence, in the concordance cosmology it takes 7.1 Gyr ($z=0.85$) for this model to double its mass. 

How is this mass distributed throughout the Galaxy? According to galaxy formation models the shape of the Galaxy is expected to vary with radius. At large radii the potential is typically dominated by a triaxial dark matter halo. In the inner-most regions the assembly of the baryonic components renders a close-to-axi-symmetric potential shape (Kazanztidis et al. 2010). The relative orientation between the principal axes of the halo and the spin vector of the Milky Way is still poorly understood. Although it is generally assumed that discs are aligned with one of the principal axes, it is also possible to find tilted configurations that are dynamically stable (e.g. Binney 1978; Vel\'azquez \& White 1999; Dubinski \& Chakrabarty 2009). To complicate this picture further, recent numerical simulations suggest that the disc-halo orientation may change repeatedly throughout the formation of spiral galaxies (e.g. Debattista et al. 2013). Therefore, in cosmologically-motivated potentials none of the components of the angular momentum may be conserved.

Here we shall bypass these theoretical uncertainties by considering a set of spherical power-law models that covers the range of potentials of astrophysical interest, i.e. with force-indices between $n=-2$ (point-mass) and $n=1$ (homogeneous density distribution), and adopting a logarithmic potential $(n=-1$) as our fiducial model for the Milky Way. This assumption allows us to concentrate on the energy evolution of tidal substructures in a growing potential, without worrying about the possible existence of other integrals of motion\footnote{Note, however, that the detection of tidal debris becomes more likely if the presence of those substructures extends to further dimensions, e.g. the angular momentum and/or metal-abundance space.}. In these potentials it can be easily shown through Equations~(\ref{eq:plaw1}),~(\ref{eq:tempentro}), and~(\ref{eq:mcosmo}) that the average temperature of tidal substructures evolves as 
\begin{equation}
\label{eq:temppl}
\frac{\mathcal{T}}{\mathcal{T}_0}=\exp\bigg[-\frac{4}{3+n}\frac{z}{1+z_c}\bigg];
\end{equation}
where $\mathcal{T}_0$ is measured at the redshift when the particles become tidally unbound from the progenitor system. 

\begin{figure}
\begin{center}
 \centerline{\includegraphics[width=84mm]{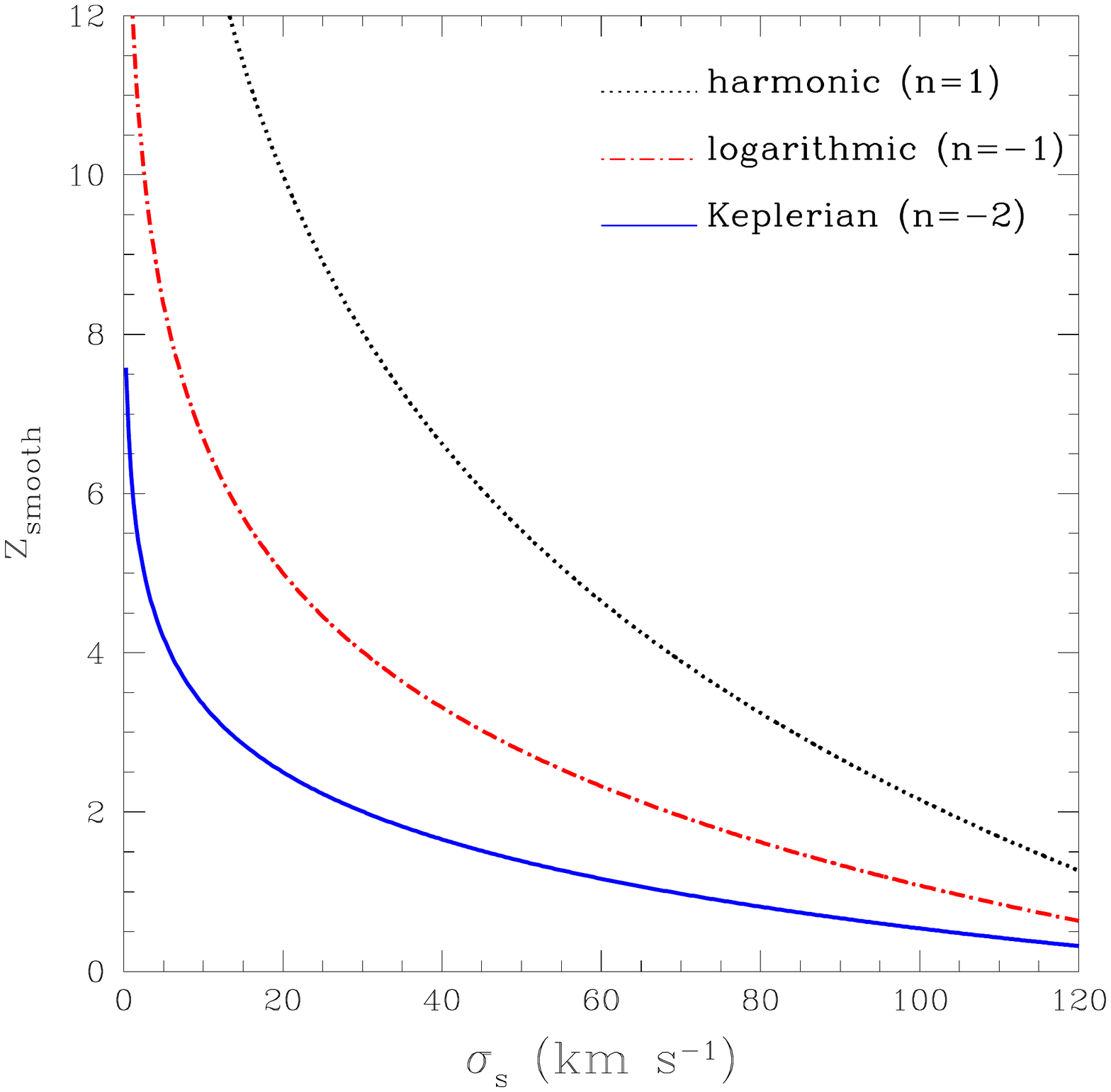}}
\vspace{-2.0truecm}
\caption{Maximum `age' of tidal fossils (measured from the redshift of merger) in the present-day configuration space as a function of the progenitor's velocity dispersion, Equation~(\ref{eq:zsmooth}). Stars that are tidally-stripped from gravitationally-bound systems at $z_{\rm age}\ll z_{\rm smooth}$ have a lower temperature at $z=0$ than the host galaxy, and may be therefore detected as substructures in the integral of motion space. In contrast, objects that were tidally disrupted at $z_{\rm age}\sim z_{\rm smooth}$ contribute to the formation of the `smooth' stellar halo. The host galaxy has a mass that grows according to Equation~(\ref{eq:mcosmo}) and a velocity dispersion $\sigma_h=220/\sqrt{2}\kms$ at $z=0$. 
Note that 'hot' substructures that originate from the disruption of massive satellite galaxies, and those orbiting in steep potentials tend to dissolve on relatively short time scales.}
\label{fig:detect}
\end{center}
\end{figure}

The detection of tidal clumps in the integral-of-motion space is limited to substructures that are currently much colder than the smooth Milky Way background, i.e. $\mathcal{T}\ll \mathcal{T}_h=(m/K_B)\sigma_h^2$; where $\sigma_h=220/\sqrt{2}\kms$ is the fiducial velocity dispersion of our Galaxy model at $z=0$. In a growing potential this condition puts a strong constrain on the maximum `age' of the substructures. Given that dynamical fossils have an energy dispersion that correlates with the dynamical mass of the progenitor system (e.g. Pe\~narrubia et al. 2006), it is useful to express temperatures in terms of mean kinetic energies. Through Equation~(\ref{eq:temppl}) the condition of detectability then becomes
\begin{equation}
\label{eq:zsmooth}
z_{\rm age}\ll z_{\rm smooth}\equiv\frac{3+n}{2}(1+z_c)\ln \bigg[\frac{\sigma_h}{\sigma_{s}}\bigg];
\end{equation}
where $\sigma_s$ is the velocity dispersion of the progenitor system at $z=z_{\rm age}$. Hence, lacking additional information (e.g. metal abundances, see Sheffield et al. 2012), the remnants of systems accreted at $z_{\rm age}\sim z_{\rm smooth}$ would be hardly distinguishable from the smooth stellar halo of the Milky Way.

Fig.~\ref{fig:detect} shows the value of $z_{\rm smooth}$ as a function of the velocity dispersion of the progenitor system. Focusing on the logarithmic potential (dashed-dotted line), which has a flat velocity curve and provides the closest representation of the Galaxy, shows that tidal debris of massive satellite galaxies such as LMC and SMC ($\sigma_s\gtrsim 80\kms$) rapidly dissolves in the stellar halo of the host. The detection of tidal debris associated to the tidal disruption of LMC-type galaxies is thus limited to the most recent ($z_{\rm age}\lesssim 0.6$) merger events. In contrast, Fig.~\ref{fig:detect} suggests that it may be possible to identify a large number of tidal clumps associated to the accretion of dwarf spheroidal galaxies ($\sigma_s\lesssim 12\kms$; see Walker et al. 2009). Tidal debris from low-mass globular clusters ($\sigma_s\lesssim 2\kms$) provide a clear-cut target for the search of substructures in the integral-of-motion space\footnote{Bear in mind, however, that the velocity dispersion of these substructures is so low that many of them may not have reached dynamical equilibrium by $z=0$. Fig.~\ref{fig:entrotime} shows that the chances of detection maximize when their orbital phase approaches apocentre.}. 

What is the impact of the Milky Way formation on the stellar halo? Dissipational processes lead to a steepening of the potential in the central regions of the Galaxy, wherein one would expect to find the largest concentration of tidal substructures. According to Fig.~\ref{fig:detect} the formation of the Galaxy will tend to accelerate the diffusion of tidal substructures that originated from early accretion events. On the other hand, the presence of a disc also enhances mass loss of satellite galaxies and stellar clusters (D'Onghia et al. 2010; Pe\~narrubia et al. 2010; Zolotov et al. 2012). Thus, the formation of the Milky Way favours the growth of both the `clumpy' and `smooth' stellar halo components.

\section{Summary and conclusions}\label{sec:conc}
This work introduces a general technique for constructing dynamical invariants (a.k.a. constants of motion) in time-dependent gravitational potentials. The method rests upon the derivation of a system of coordinates in which the explicit time-dependence is removed from the Hamiltonian. After carrying out the inverse transformation the integrals of motion admitted by the gravitational potential become dynamical invariants in the original coordinates. 

By construction, dynamical invariants are conserved quantities along the phase-space path of a particle motion. In practical terms this means that the differential equations that define the coordinate transformation and those that determine the motion of particles through phase-space are coupled. 
However, in a few exceptional cases both sets of equations can be de-coupled, thus allowing the derivation of {\it exact} invariants. This is the case, for example, of the harmonic potential (see Feix et al. 1987) as well as Dirac's cosmology, where Newton's constant $G$ varies as the reciprocal of the time (Lynden-Bell 1982). In a regime where the mean field varies slowly it is possible to derive approximate invariants for power-law forces, $F(\xi,t)=-\mu(t)\xi^n$, where $\epsilon \equiv \dot \mu/\mu_0 \lesssim T_0^{-1}$, and $T_0$ is the radial period of an orbit. Numerical tests show that these quantities are conserved at order $|\Delta I /I_0|\lesssim 0.1 (\epsilon T_0)^{2}$ for the range of power-law forces of astronomical interest ($-2\le n \le 1$).

This technique offers advantages over standard perturbation methods. 
For example, while actions are only conserved in systems that evolve adiabatically, dynamical invariants stay constant independently of the time scale for change in the potential. Except for a few rare cases that admit exact invariants (see above), the construction of {\it analytical} invariants is only possible for scale-free potentials that vary slowly. However, it is worth noting that approximate invariants remain accurate even outside the adiabatic regime ($\epsilon T_0\lesssim 1$), as shown in Fig.~\ref{fig:eps}. In general, for scaled potentials the transformation $R(t)$ is coupled to the trajectory in phase-space of individual particles and needs to be computed numerically.

The derivation of dynamical invariants yields tight constraints on the dynamical evolution of collisionless systems. For example, invariants can be used to describe the evolution of the microcanonical distribution of gravitating systems without relying on ergodicity or probability assumptions. As an illustration, we consider the case of tidal streams orbiting in a logarithmic potential whose circular velocity can either grow or drop linearly with time. Restricted $N$-body simulations show that tidal tails exhibit fluctuations in entropy, temperature and specific heat that damp out as these systems approach dynamical equilibrium. This behaviour can
not be described by the canonical distribution, which evolves toward a suitable equilibrium configuration through maximization of entropy (e.g. Penrose 1979). For gravitating systems this is equivalent to maximizing the phase-space volume available to the particle ensemble (Padmanabhan 1990). However, substructures that have not mixed in phase space violate this condition, as the distribution function oscillates in phase with the radial motion of their orbits. These systems also violate the ergodic hypothesis, which assumes that the time-averaged properties of microcanonical ensembles can be derived from a phase-space average over all possible microstates. In contrast, dynamical invariants allow us to describe the statistical properties of tidal tails through a simple time averaging of deterministic equations. We show that the equivalence between the micro and macrocanonical descriptions only emerges as tidal tails progressively fill the phase-space volume available to their orbits and a state of dynamical equilibrium is reached. 

Merger substructures tend to {\it diffuse} in the integral-of-motion space throughout the growth of the host potential. In galaxies that build up mass hierarchically, a {\it smooth} stellar halo emerges as the inescapable by-product of the deposition and progressive dissolution of dynamical fossils. Given the stochasticity of merger trees, substructures in the stellar halo are expected to cover a continuous spectrum of temperatures. Attempts to quantify the amount of substructure in the stellar halo of our Galaxy (e.g. Bell et al. 2008; Starkenburg et a. 2009; Schlaufman et al. 2010; Xue et al. 2011) are biased toward the coldest and youngest substructures and must therefore be taken as lower limits. For example, using cosmologically-motivated models we estimate that the detection of tidal debris associated to massive satellites (i.e. LMC-type galaxies) is limited to the most recent events, $z_{\rm age}\lesssim 0.6$, in gross agreement with the results derived from $N$-body models of structure formation (e.g. Font et al. 2008; Johnston et al. 2008). This suggests that the majority of substructures identifiable as dynamical fossils in the present-day configuration space likely originate from the tidal stripping of low-mass objects, such as dwarf spheroidals and stellar clusters. A noteworthy remark refers to the active role that baryons may play in the formation of stellar haloes. Dissipational processes in the host galaxy accelerate both the disruption rate of gravitationally-bound objects and the `dissolution' of tidal substructures through a steepening of the central potential. 

Further applications of dynamical invariants to gravitating systems approaching an equilibrium state will be explored in separate contributions. 

\section{Acknowledgements}
This work has greatly benefited from the comments and suggestions of Douglas Heggie, John Peacock, Andrew Pontzen and Matt Walker. The generous input of James Binney regarding the analysis of actions is appreciated. Also, a word of thanks to the anonymous referee for his/her very useful comments. 

{}

\end{document}